\documentclass{article}
\usepackage{wds10,epsf,rotating,multirow}
\usepackage[square]{natbib}
\lefthead{TORBANIUK ET AL.}
\righthead{SOME SPECTRAL PROPERTIES OF THE QUASAR UV BUMP}

\begin{document}

\title{Some spectral properties of the quasar ultraviolet bump}

\author{O. Torbaniuk}
\affil{Taras Shevchenko National University of Kyiv, Faculty of Physics, Kyiv, Ukraine.}
\author{G. Ivashchenko}
\affil{Taras Shevchenko National University of Kyiv, Astronomical Observatory, Kyiv, Ukraine.}
\author{O. Sergijenko}
\affil{Ivan Franko National University of Lviv, Astronomical Observatory, Lviv, Ukraine.}

\begin{abstract}
In the present work the part of the quasar UV-optical bump within the wavelength range $1210-1450$\,\AA\ was studied with the help of composite spectra compiled from the samples of SDSS DR7 spectra with the similar spectral index $\alpha_{\lambda}$ within $1270-1480$\,\AA. This division allowed to see weak emission lines, which were not detected in previous studies of the quasar composite spectra, but were known from individual optical or composite UV spectra from the Hubble Space Telescope. Although the physical explanation of the difference in spectral indices between quasars and their dependence on quasar parameters is still not clear, it is obvious that this difference has to be taken into account when generating composite spectra, e.\,g. for redshift measurements. It was also shown that the equivalent width of the emission lines does not depend on the spectral index. 
\end{abstract}

\begin{article}

\section{Introduction}

The spectral energy distribution of quasars in UV-optical range is characterised by the so called Big Blue Bump with a pick around $\lambda_{rest}=1000-1300$\,\AA, broad emission lines of hydrogen Lyman and Balmer series and low-ionized metals (e.\,g. C{\sc iv}, Si{\sc iii}, Mg{\sc ii} etc.), broad absorption lines (only in $\sim10$\% of quasars) and decrement of the flux blueward of 1215\,\AA\ due to absorption by intergalactic H{\sc i} (the Ly$\alpha$ forest). It is believed that the (mostly thermal) continuum and emission lines originate from the hot accretion disk and circumnuclear fast moving clumps of gas, correspondingly. The proximity of these regions is considered to be the most promising explanation of the Baldwin effect \citep{baldwin}: the inverse correlation of equivalent width of some emission lines with the monochromatic fluxes at UV region (in most cases the flux at $1450$\,\AA\ or $2500$\,\AA\ is considered in the literature \citep[e.\,g.][]{dietrich+2002,wu+2009}).  

Due to remarkable similarity of the quasar spectra from object to object, the mean, or composite, spectra are usually used to study the general spectral properties of quasars. They are also widely used as templates for redshift determination (mainly in automatic redshift surveys). The averaging over several hundreds or even thousands of (usually) medium-resolution spectra allows to increase the signal-to-noise ratio and to reveal tiny features unseen in individual spectra. The composite spectra of quasars in ultraviolet-optical band were compiled for a wide set of quasar samples \citep{brotherton+01,francis+91,scott2+04,vandenberk+01}. They also used in Ly$\alpha$-forest studies \citep[e.\,g.][]{bernardi+03,desjacques+07,polinovskyi+10}. One of the main sources of uncertainties during the generation of the quasar composite spectra is the difference in spectral shape of individual objects (despite of the general similarity), e.\,g. the difference in spectral index $\alpha_{\lambda}$ of the red part of UV-
optical bump. Thus in the present work to study the part of the UV-optical bump we used the composite spectra compiled by us from samples of spectra with similar $\alpha_{\lambda}$.

\section{The data}

For our study we used 16 composite spectra of quasars compiled by us \citep{ivashchenko}. Each of these spectra was generated of 200 individual ones with similar spectral indices $\alpha_{\lambda}$ within the range $1270-1480$\,\AA\, namely from those that have the values of $\alpha_{\lambda}$ close to $-0.6-k\cdot0.1$, where $k=1..16$ is the composite number. These subsamples were taken from the publicly available release of the sky-residual subtracted spectra \citep{wild_10} for the Sloan Digital Sky Survey (SDSS) Legacy Release 7 \citep{Abazadjian_2009}. The spectral indices for individual spectra were calculated considering the continuum of the quasar spectra redward of the Ly$\alpha$ emission line (1215\,\AA) to be a power-law $\sim\lambda^{\alpha_{\lambda}}$, and using the following wavelength ranges, which are the most free from emission lines: 1278$-$1286, 1320$-$1326, 1345$-$1360 and 1440$-$1480\,\AA\AA. The mean arithmetic spectra were obtained after (i) division of each individual spectrum on its 
normalization constant (the mean flux over all pixels within the rest wavelength range 1450--1470\,\AA), (ii) rebinning them with $\Delta\lambda_{rest}=2$\,\AA, and (iii) stacking the spectra into the rest frame. The dispersion $\sigma^{2}$ of each pixel of the composite spectrum is calculated from the noises $\sigma_{i}$ of pixels of individual spectra as $\sigma^{-2}=\sum\limits_{i}\sigma_{i}^{-2}$.

\begin{figure}[htb]
\begin{center}
\begin{tabular}{c}
\epsfxsize=.7\linewidth
\epsfbox{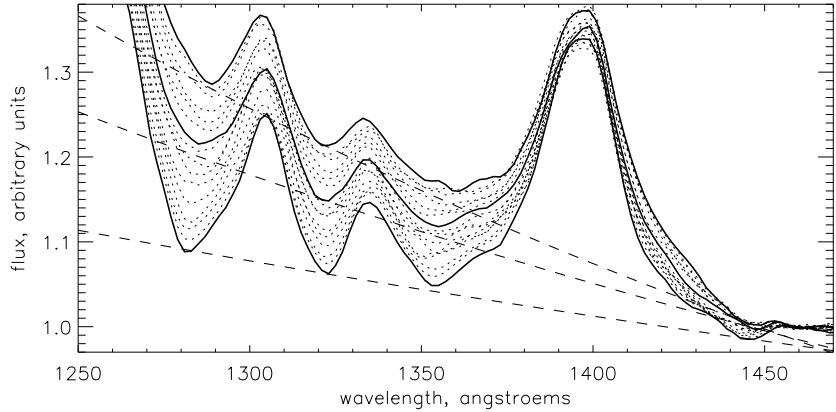}
\end{tabular}
\end{center}
\caption{The parts of the composite spectra used for continuum approximation. See explanation in the text.}\label{fig:lena}	
\end{figure}

The spectral indices of the composite spectra were calculated in the way mentioned above, but the parts of the spectra which are the most free from emission lines were selected manually in each composite. The parts of spectra used for calculating the spectral index are shown in Figure~\ref{fig:lena} (\textit{short-dashed} line, the highest spectra are the steepest ones). Three spectra with the slopes $-0.84$, $-1.49$ and $-2.12$ are shown with the \textit{solid} lines together with the fitted power-law continua (\textit{dashed} lines). The regions used for continuum fitting slightly vary from spectrum to spectrum which is probably caused by changing of intensity of weak emission lines. It is worth to note that there is no absolutely emission line-free region in UV-optical bump, and selection of lines-free regions means selection of regions with the weakest emission lines.

\vspace*{-3ex}
\begin{table*}[!h]
\begin{center}
\caption{The mean redshift $\bar{z}$, spectral index $\alpha_{\lambda}$, and the mean monochromatic luminosity $\langle \log l_{1450}\rangle$ at 1450\,\AA\ for 16 composites.}\label{tab:samples}
\begin{tabular}{c|c|c|c}
\hline
n & $\alpha_{\lambda}$ & $\bar{z}$ & $\langle\log l_{1450}\rangle$ \\
\hline
  1 & $-0.84\pm0.04$ & $2.90\pm0.54$ & $42.70\pm0.21$ \\
  2 & $-0.92\pm0.03$ & $2.89\pm0.53$ & $42.72\pm0.21$ \\
  3 & $-0.99\pm0.02$ & $2.87\pm0.51$ & $42.72\pm0.21$ \\
  4 & $-1.02\pm0.05$ & $2.85\pm0.49$ & $42.72\pm0.24$ \\
  5 & $-1.19\pm0.02$ & $2.82\pm0.43$ & $42.73\pm0.24$ \\
  6 & $-1.30\pm0.02$ & $2.84\pm0.49$ & $42.74\pm0.21$ \\
  7 & $-1.42\pm0.03$ & $2.82\pm0.44$ & $42.76\pm0.23$ \\
  8 & $-1.42\pm0.02$ & $2.76\pm0.40$ & $42.78\pm0.21$ \\
  9 & $-1.49\pm0.05$ & $2.77\pm0.44$ & $42.78\pm0.25$ \\
  10 & $-1.55\pm0.03$ & $2.80\pm0.46$ & $42.79\pm0.27$ \\
  11 & $-1.67\pm0.01$ & $2.75\pm0.45$ & $42.78\pm0.25$ \\
  12 & $-1.83\pm0.04$ & $2.75\pm0.43$ & $42.78\pm0.22$ \\
  13 & $-1.86\pm0.04$ & $2.74\pm0.43$ & $42.77\pm0.24$ \\
  14 & $-1.99\pm0.03$ & $2.72\pm0.42$ & $42.74\pm0.23$ \\
  15 & $-2.00\pm0.02$ & $2.74\pm0.42$ & $42.74\pm0.23$ \\
  16 & $-2.12\pm0.03$ & $2.74\pm0.43$ & $42.75\pm0.24$ \\
\hline
\end{tabular}
\end{center}
\end{table*}

The spectral indices for the composite spectra, the mean redshift of the corresponding sample, and the mean logarithm of the quasar monochromatic luminosity at $1450$\,\AA, $\langle\log l_{1450}\rangle$, are presented in Table~1. The the value $\log l_{1450}$ for individual spectra was calculated using the mean flux within the wavelength range $1449-1451$\,\AA\ and photometric distance to the quasar obtained within the spatially flat $\Lambda$CDM cosmological model with the matter density parameter $\Omega_{M}=0.27$ and Hubble constant $H_{0}=70.5$\,km\,s$^{-1}$\,Mpc$^{-1}$. The quoted error bars are the root mean squares for the corresponding distributions for  $z$ and $\langle\log l_{1450}\rangle$, and 1$\sigma$ errors for $\alpha_{\lambda}$, correspondingly. 

\section{The method}

\begin{figure}[!h]
\begin{center}
\begin{tabular}{c}
\epsfxsize=.4\linewidth
\epsfbox{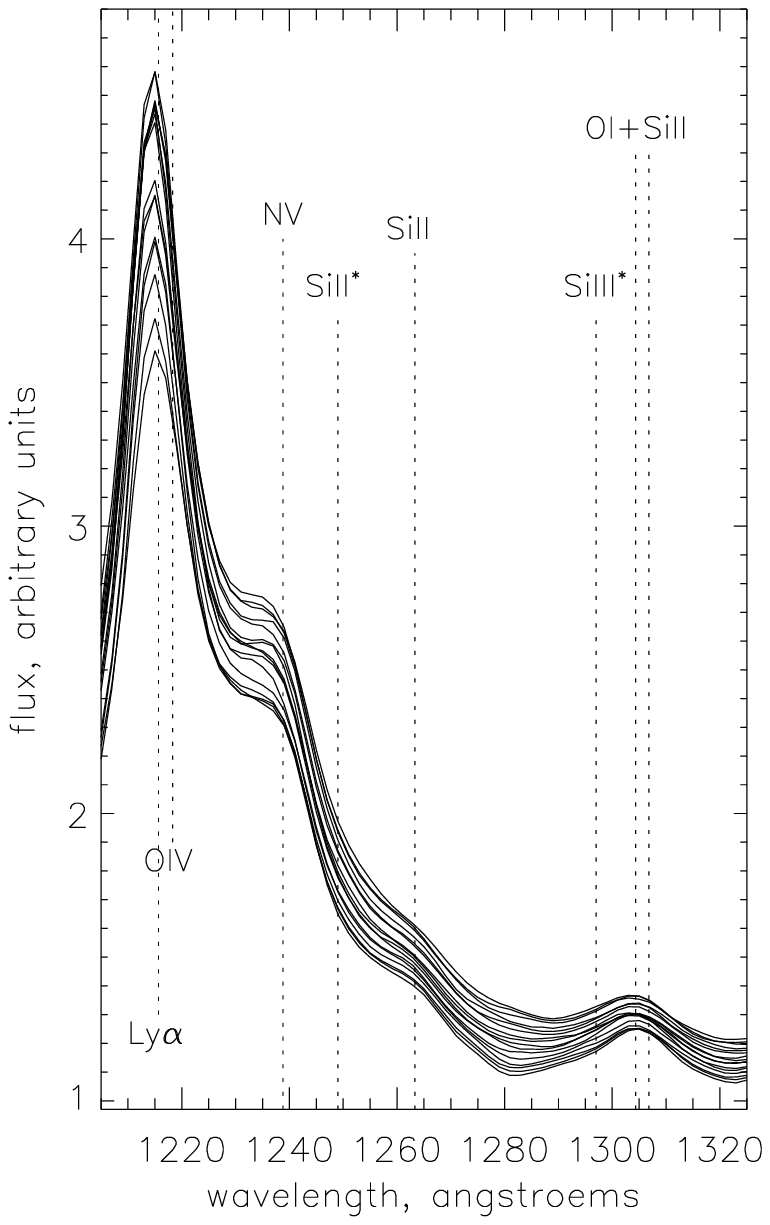}
\hfill
\epsfxsize=.4\linewidth
\epsfbox{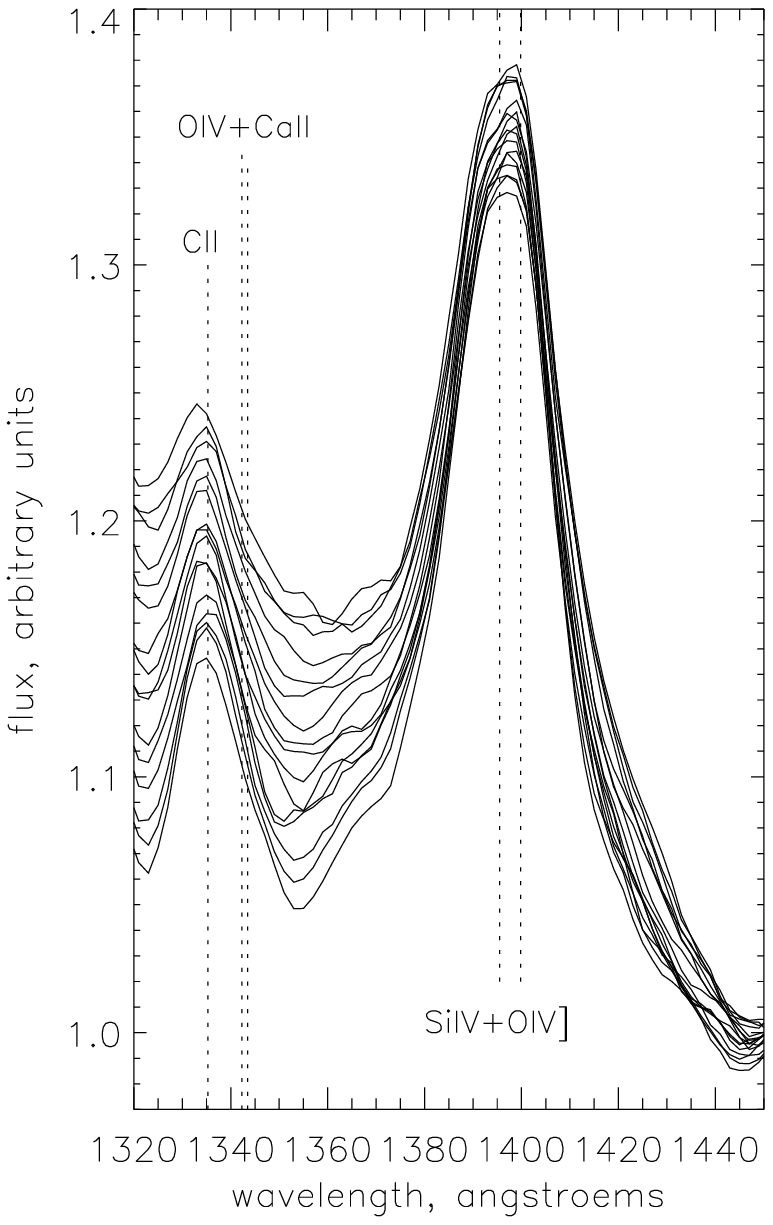}
\end{tabular}
\vspace*{-4ex}
\caption{Wavelength ranges 1215--1320~\AA\ (left) and 1323--1443~\AA\ (right). Dashed lines indicate the central wavelengths of the emission lines found by \citet{vandenberk+01}. Here asterisk indicates more than one allowed value of the total angular momentum J for that specific term and transition, ``]'' means  intercombination transitions. }
\label{fig:spec_m1234}
\end{center}
\end{figure}

The wavelength ranges (a) $1215 - 1321$\,\AA\ and (b) $1323-1447$\,\AA\ (Figure~\ref{fig:spec_m1234}) were considered separately and fitted with a superposition of continuum and the smallest possible number of emission lines in the form:
\begin{equation}\label{eq:model}
f(\lambda^{rest})=C+\sum\limits_{i}a_{i}\exp\left[-\frac{(\lambda^{rest}-\lambda^{0}_{i})^{2}}{2w_{i}^{2}}\right], 
\end{equation}
where in case (a) $C=b$, i.\,e. the continuum is fitted with a constant, and in case (b) $C=d\lambda^{\alpha_{\lambda}}$ with fixed $\alpha_{\lambda}$ from Table~1. The values of $b$, $d$, $a_i$, $w_{i}$, $\lambda^{0}_{i}$ were calculated in the following two steps: (i) using the \texttt{IDL lmfit} subroutine for each multiplet the best fit model (based on the minimal $\chi^2$/d.o.f.) in a form of (\ref{eq:model}) was found; (ii) the central wavelengths $\lambda^{0}$ obtained at the first step were fixed and the best fit values of other parameters $\{b,d,a_{k},w_{k}\}$ with the 1$\sigma$ errors from the extremal values of the N-dimensional distribution were calculated by the Markov Chain Monte Carlo (MCMC) method using \texttt{CosmoMC} package \citep{cosmomc} as a generic sampler (with the values of $\{b,d,a_{k},w_{k}\}$ obtained at the first step as starting values). The MCMC technique has been chosen as it is a fast and accurate method of exploration of high-dimensional parameter spaces. In each case we 
generated 8 chains which have converged to $R-1<0.0075$. Due to the small values of the flux dispersion $\sigma_{f}^{2}$ in composite spectra determined from the covariance matrix to estimate errors correctly we introduced the additional intrinsic errors $\sigma_{int}$ such that the total dispersion, $\sigma^2 = \sigma_f^2+\sigma_{int}^2$, results in the minimal $\chi^2$/d.o.f. to be $1-0$.

For separate lines or sum of lines the equivalent widths were calculated. In each case the area under the lines was calculated as an integral of an analytic function (Gaussian or several Gaussians) with the obtained parameters.

\section{Results and discussion}

\begin{figure}[htb]
\begin{center}
\begin{tabular}{c}
\epsfxsize=.75\linewidth
\epsfbox{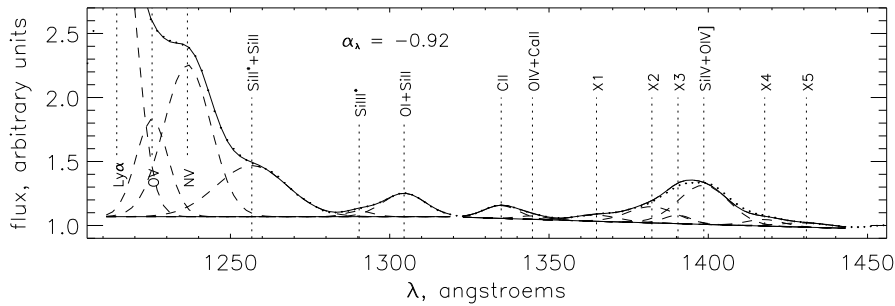}
\vspace*{-4ex}
\end{tabular}
\end{center}
\caption{The composite spectrum with $\alpha_{\lambda}=-0.92$ with the best fits for both two regions. The dashed lines indicate the found emission lines, while short-dashed lines indicate their central wavelengths. X1 -- X5 stand for lines which were not identified by us.}\label{fig:example}	
\end{figure}

In Figure~\ref{fig:example} one of composite spectra with $\alpha_{\lambda}=-0.92$ is shown with the best fits for both regions as an example. The dashed lines indicate the emission lines found in spectrum. The labels X$_{1}$-X$_{5}$ stand for components, identification of which was not found in the literature. In Table~\ref{table:emission_lines} the central wavelengths of emission lines from all 16 spectra are presented along with possible identification and corresponding laboratory rest wavelengths. The obtained parameters for each line (amplitude and FWHM) will be presented elsewhere. The identification of lines was made using known emission lines from the literature. The following lines were found by \citet{vandenberk+01} in composite spectrum of 2200 quasars from SDSS without taking into account any differences in spectra shape: Ly$\alpha$+O{\sc iv} (one feature), N{\sc v}, Si{\sc ii}, O{\sc i}+Si{\sc ii} (one feature), C{\sc ii} and Si{\sc iv}+O{\sc iv]} (one feature). The separate components of these 
emission features as far as other lines found in the present work are known from works on ultraviolet individual \citep{laor+1994,laor+1995,laor+1997} and composite \citep{brotherton+1994} spectra of quasars from the Hubble Space Telescope, and from empirical UV template for Fe emission in quasars obtained by \citet{vestergaard} from the galaxy I~Zwicky~1. All the lines, for which we did not find any possible candidates in the literature could be new lines, not noticed by other authors due to their small intensity, as well as additional components of the known lines originated due to some special physical conditions in the broad line regions in a quasar. 

\begin{figure}[!h]
\begin{center}
\begin{tabular}{c}
\epsfxsize=.4\linewidth
\epsfbox{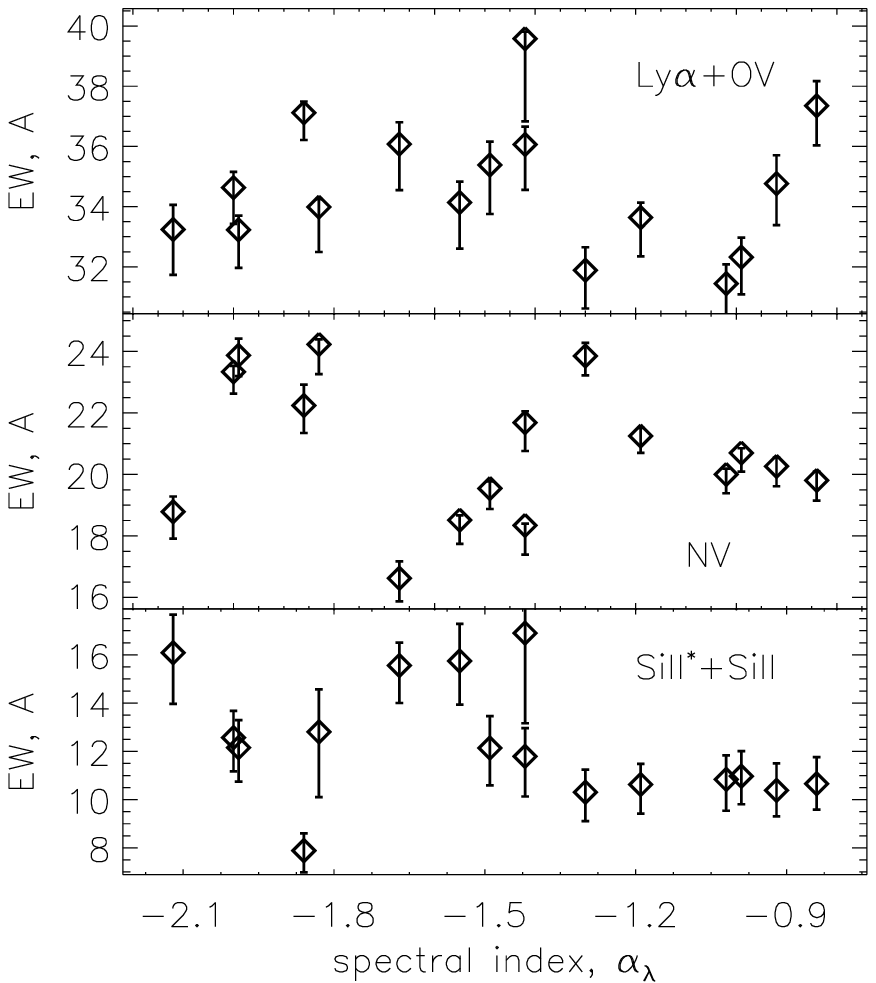}
\hfill
\epsfxsize=.4\linewidth
\epsfbox{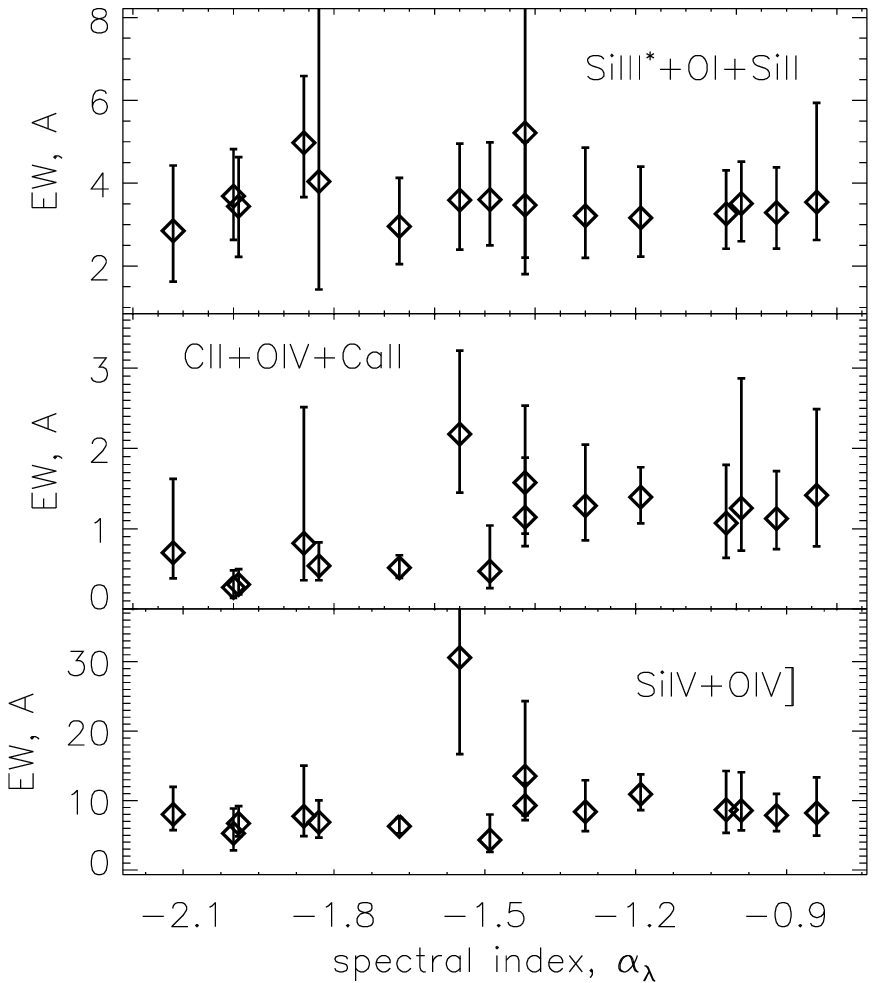}
\end{tabular}
\caption{The spectral index -- equivalent width of lines or sum of lines (in \AA) diagram.}
\label{fig:ew}
\end{center}
\end{figure}

In Figure~\ref{fig:ew} the spectral index -- equivalent width (for separate lines or sum of lines) diagrams are presented. We applied the F-test and found that for all emission features presented in this figure the values of equivalent width do not depend on the spectral index. This result is an expected one if we take into account the Baldwin effect and the absence of luminosity-dependence of the spectral index, as it is seen from Table~\ref{tab:samples}. On the other hand, the mean redshift of the subsamples slightly increases with $\alpha_{\lambda}$, that could be an evidence for evolution of $\alpha_{\lambda}$ with time, but also can be caused by some  selection effects. Although the physical explanation of the difference in spectral indices between quasars and their dependence on quasar parameters is still not clear, it is obvious that this difference has to be taken into account when generating the composite spectra, e.\,g. for redshift determination. When using composite spectra made without taking 
into 
account the difference in spectral shape one usually deals with larger noise and thus treats blended emission features as one line, increasing the errors in central wavelengths and, hence, in line identification.  


\acknowledgments 
{This work has been supported by Swiss National Science Foundation  (SCOPES grant No 128040). The authors are thankful  to the Sloan Digital Sky Survey team. Funding for the SDSS has been provided by the Alfred P. Sloan Foundation, the Participating Institutions, the National Aeronautics and Space Administration, the National Science Foundation, the US Department of Energy, the Japanese Monbukagakusho, and the Max Planck Society. The authors also acknowledge the usage of \texttt{CosmoMC} package.}
\nopagebreak

\begin{sidewaystable*}[!h]
\centering
  \caption{The cental wavelength of emission lines found in spectra. Lines in brackets are probably components of one found emission feature. X1 etc. stand for lines which were not identified by us. Here asterisk indicates more than one allowed value of the total angular momentum J for that specific term and transition, ``]'' means  intercombination transitions.} \label{table:emission_lines}
\vspace*{2ex}
\fontsize{9}{12}\selectfont
\centering
\begin{tabular}{|c|c|c|c|c|c|c|}
\hline
  & 1 & 2 & 3 & 4 & 5 & 6 \\
\hline
  & Ly$\alpha+$O{\sc iv} & N{\sc v} & Si{\sc ii}$^\ast+$Si{\sc ii} & Si{\sc iii}$^\ast+$O{\sc i}$+$Si{\sc ii} & C{\sc ii}$+$O{\sc iv}$+$Ca{\sc ii} & Si{\sc iv}$+$O{\sc iv}$]$+X$_{1}$+\ldots \\
\hline
  \multirow{3}{*}{$\lambda_{lab}$} & \multirow{3}{*}{1215.7+1218.3} &  & [1248.4+1251.1]+ & [1295.5+1298.9]+ & \multirow{3}{*}{1335.3} & \\
    &  & {1238.8+} & +[1260.4+1264.7+ & +[1302.2+1304.6+ &  & 1395.5+1399.8+  \\
    &  & {+1242.8} & +1265.0] & +1306.0+1309.3] &  & +1401.8+1407.4\\
 \hline
1 & 1214.4$+$1225.6 & 1236.9 & 1256.7 & 1290.4$+$1304.4 & 1334.5$+$1346.7 & 1368.2$+$1382.5$+$1390.4$+$1399.8$+$1418.7$+$1436.3 \\
2 & 1214.3$+$1225.4 & 1236.5 & 1256.7 & 1290.4$+$1304.5 & 1335.0$+$1344.8 & 1364.9$+$1382.3$+$1390.5$+$1399.6$+$1417.7$+$1430.8 \\
3 & 1214.3$+$1225.0 & 1236.1 & 1256.7 & 1289.7$+$1304.1 & 1334.9$+$1343.2 & 1364.4$+$1380.1$+$1390.3$+$1399.3$+$1418.7$+$1431.4 \\
4 & 1214.8$+$1226.4 & 1236.4 & 1257.0 & 1290.6$+$1304.2 & 1334.7$+$1341.4 & 1364.8$+$1376.9$+$1386.5$+$1392.0$+$1399.0$+$1412.3$+$1419.6$+$1439.5 \\
5 & 1214.3$+$1225.3 & 1236.0 & 1256.7 & 1291.4$+$1304.5 & 1335.8 & 1370.5$+$1396.5$+$1401.7$+$1410.9$+$1422.4$+$1436.4 \\
6 & 1214.2$+$1224.3 & 1234.8 & 1256.5 & 1290.1$+$1304.4 & 1334.6$+$1348.5 & 1361.4$+$1382.1$+$1389.8$+$1399.3$+$1407.4$+$1435.4 \\
7 & 1213.2$+$1221.6 & 1236.1 & 1253.4 & 1300.4$+$1305.2 & 1334.9 & 1362.6$+$1394.5$+$1396.5$+$1418.9 \\
8 & 1213.8$+$1223.7 & 1235.9 & 1255.7 & 1285.7$+$1303.6 & 1334.2$+$1344.1 & 1363.0$+$1382.0$+$1390.3$+$1399.6$+$1413.3$+$1430.7$+$1438.6 \\
9 & 1213.4$+$1223.4 & 1236.3 & 1255.6 & 1286.6$+$1303.6 & 1334.1$+$1343.2 & 1378.0$+$1394.0$+$1400.2$+$1405.8$+$1415.7$+$1425.6$+$1438.3 \\
10 & 1213.8$+$1223.9 & 1235.5 & 1252.7 & 1288.6$+$1304.6 & 1332.1 & 1372.3$+$1395.4$+$1400.6$+$1407.9$+$1418.5$+$1428.4 \\
11 & 1213.7$+$1223.4 & 1236.2 & 1252.3 & 1288.4$+$1304.0 & 1334.0$+$1342.3 & 1376.2$+$1395.6$+$1401.5$+$1408.7$+$1421.0$+$1436.6 \\
12 & 1214.5$+$1225.7 & 1235.1 & 1256.0 & 1281.9$+$1301.6$+$1304.9 & 1334.8$+$1345.1 & 1385.1$+$1393.9$+$1401.0$+$1407.8$+$1423.3$+$1438.5 \\
13 & 1214.5$+$1225.8 & 1237.4 & 1250.8$+$1258.4 & 1276.7$+$1304.5 & 1333.9$+$1336.0 & 1376.0$+$1397.9$+$1419.3$+$1429.8 \\
14 & 1214.4$+$1225.0 & 1235.1 & 1255.7 & 1286.2$+$1304.0 & 1335.3 & 1388.8$+$1398.5$+$1422.1 \\
15 & 1213.7$+$1223.2 & 1235.5 & 1256.1 & 1283.7$+$1303.2 & 1335.8 & 1396.8$+$1397.3 \\
16 & 1214.1$+$1224.1 & 1235.6 & 1252.4 & 1285.6$+$1302.2 & 1333.7$+$1343.7 & 1371.8$+$1397.4$+$1416.8$+$1425.7 \\
\hline
\end{tabular}
\end{sidewaystable*}
\end{article}
\end{document}